\documentclass[lettersize,journal]{IEEEtran}
\usepackage{amsmath,amsfonts}
\usepackage{algorithmic}
\usepackage{array}
\usepackage[caption=false,font=normalsize,labelfont=sf,textfont=sf]{subfig}
\usepackage{textcomp}
\usepackage{stfloats}
\usepackage{url}
\usepackage{verbatim}
\usepackage{graphicx}
\def\BibTeX{{\rm B\kern-.05em{\sc i\kern-.025em b}\kern-.08em
    T\kern-.1667em\lower.7ex\hbox{E}\kern-.125emX}}
\usepackage{balance}
\usepackage{caption}
\usepackage{pifont}
\usepackage{textcomp}
\usepackage{blindtext}
\usepackage{booktabs}
\usepackage{lipsum}
\usepackage{soul}
\usepackage{adjustbox}
\usepackage[normalem]{ulem}
\useunder{\uline}{\ul}{}
\usepackage [inline]{enumitem}
\usepackage{adjustbox,lipsum}
\usepackage{booktabs}
\usepackage[table]{xcolor}
\usepackage[normalem]{ulem}
\usepackage[utf8]{inputenc}
\usepackage[english]{babel}
\usepackage{dirtytalk}
\usepackage{longtable}            
\usepackage{rotating}
\usepackage{multirow}
\usepackage{amsmath} 
\usepackage{hyperref}
\usepackage{comment}
\usepackage{url}
\usepackage{amssymb}
\usepackage{pifont}
\usepackage{xcolor}
\usepackage[font=small]{caption}
\usepackage{svg}

\usepackage{scalerel}
\usepackage{subfig}
\usepackage[font=normal,labelfont=normal]{caption}
\usepackage{threeparttable}
\usepackage{makecell} 

\usepackage{subcaption} 

\begin{document}

\title{Trustworthy Inter-Provider Agreements in 6G Using a Privacy-Enabled Hybrid Blockchain Framework}

\author{
    \IEEEauthorblockN{Farhana Javed and Josep Mangues-Bafalluy}\\
    \IEEEauthorblockA{
        \textit{Services as networkS (SaS), Centre Tecnològic de Telecomunicacions de Catalunya (CTTC/CERCA), Castelldefels, Spain}\\
        Email: \{farhana.javed, josep.mangues\}@cttc.es
    }
    \thanks{This work was partially supported by MCIN/AEI grant PID2021-126431OB-I00, “ERDF A way of making Europe”, Spanish MINECO grants TSI-063000-2021-54/-55 (6G-DAWN), and Generalitat de Catalunya grant 2021 SGR 00770.}
}



\maketitle

\begin{abstract}
Inter-provider agreements are central to 6G networks, where administrative domains must securely and dynamically share services. To address the dual need for transparency and confidentiality, we propose a privacy-enabled hybrid blockchain setup using Hyperledger Besu, integrating both public and private transaction workflows. The system enables decentralized service registration, selection, and SLA breach reporting through role-based smart contracts and privacy groups. We design and deploy a proof-of-concept implementation, evaluating performance using end-to-end latency as a key metric within privacy groups. Results show that public interactions maintain stable latency, while private transactions incur additional overhead due to off-chain coordination. The block production rate governed by IBFT 2.0 had limited impact on private transaction latency, due to encryption and peer synchronization. Lessons learned highlight design considerations for smart contract structure, validator management, and scalability patterns suitable for dynamic inter-domain collaboration. Our findings offer practical insights for deploying trustworthy agreement systems in 6G networks using privacy-enabled hybrid blockchains.

\end{abstract}

\begin{IEEEkeywords}
Blockchain, DLT, smart contracts, 6G, inter-provider , Ethereum, Hyperledger Besu.
\end{IEEEkeywords}

\section{Introduction}
\label{sec1}

The next generation of mobile networks (6G) envisions large-scale service delivery across multiple administrative domains, requiring dynamic, trust-driven coordination. Network Function Virtualization (NFV) plays a foundational role in this vision, enabling infrastructure to be flexibly decomposed and recomposed into service slices spanning providers. This shift supports new service delivery models—such as those promoted by CAMARA \cite{camara} and GSMA’s Open Gateway framework—by allowing providers to offer, consume, and broker network functions through open interfaces.

However, this openness increases operational complexity and introduces trust challenges. When service providers collaborate—for example, during surges in demand at large-scale events—they require agile mechanisms to lease and exchange resources across domains. Traditional centralized trust models are insufficient for such decentralized arrangements, prompting the need for distributed approaches to secure transactions and enforce contracts \cite{ETSIGSPDL024,3GPPTS23.251,ETSI2020Applications}. Recognizing this, initiatives like NGMN emphasize "trustworthiness by design" in 6G, explicitly addressing transparency, privacy, reliability, and safety, and identifying blockchain as a candidate technology for multi-party coordination \cite{NGMN2023}.

Blockchain and Distributed Ledger Technologies (DLTs) offer tamper-evident, decentralized record-keeping and have been actively explored within the telecommunications sector. Industry efforts—such as TM Forum’s Catalyst Program, ETSI’s Permissioned Distributed Ledgers initiative, and CAMARA’s integration of telecom identifiers with Web3 services—demonstrate blockchain’s potential for automating settlement, Quality of Service (QoS) monitoring, Service Level Agreement (SLA) enforcement, and decentralized identity management \cite{TMForumCatalystExample,ETSI2019PDLInterop,ETSI2021SmartContracts}. These examples illustrate a broader trend toward enabling transparent, multi-vendor coordination with reduced operational overhead.

Despite significant exploration, most blockchain implementations in telecom remain conceptual or assume uniform access models. Academic research similarly emphasizes high-level architectures for privacy, resource sharing, and multi-domain orchestration in 5G and beyond \cite{faisal2022beat, antevski2022federation, xevgenis2025blockchain}, but lacks detailed analysis of transaction-level behavior, especially in hybrid public-private execution paths. Existing studies on scalability \cite{zahir2024performance, wilhelmi2022end}, economic trade-offs \cite{afraz2023blockchain}, and consensus performance \cite{antevski2023applying, suetor2025overview} focus largely on block-level optimization or multi-cloud architectures rather than privacy-enabled, transaction-level dynamics. Prior work, including our own \cite{javed2022blockchainglobecom, JavedNFVworkshop, javed2025empiricalsmartcontractslatency}, has investigated latency and gas costs in EVM environments but has not examined how private transactions impact contract structure, performance, and network behavior. To our knowledge, no existing work provides an implementation-driven analysis of modular smart contracts within a privacy-enabled permissioned blockchain tailored for inter-provider agreements in 6G.

In this paper, we address this gap by designing and implementing a privacy-enabled hybrid transaction framework for inter-provider agreements in 6G networks. Our system is built on Hyperledger Besu, a permissioned blockchain platform with native support for private transactions via Tessera. The architecture enables both public service discovery and confidential contract enforcement, using privacy groups to isolate sensitive information—such as SLA violations—while allowing open access to shared service metadata. To support modularity and role separation, we implement a set of role-based smart contracts, each responsible for distinct agreement phases. The framework aligns with the 5Growth NFV-MANO architecture, ensuring integration compatibility with existing telecom orchestration systems. Beyond system design, we deploy a multi-node testbed and conduct an empirical evaluation under realistic conditions, measuring end-to-end latency across public and private transaction workflows to assess the impact of privacy enforcement on execution performance.

Our key contributions are as follows: \begin{itemize} \item We propose a hybrid blockchain system combining public transparency with private data isolation to support inter-provider agreement enforcement in 6G. \item We implement modular smart contracts for service registration, selection, and SLA breach management, mapping public and private functions to appropriate blockchain layers via Besu and Tessera. \item We empirically evaluate public and private transaction behavior using end-to-end latency as a metric, quantifying the overhead introduced by encrypted communication, peer coordination, and off-chain payload handling. \item We extract actionable design lessons for privacy-enabled DApps in 6G, including contract modularization, privacy group reuse, role-based credential handling, and validator configuration strategies under IBFT 2.0. \end{itemize}

The remainder of this article is structured as follows: Section~\ref{sec2} reviews related work and outlines privacy challenges in inter-provider blockchain systems. Section~\ref{sec3} describes the proposed architecture, key components, and smart contract design. Section~\ref{sec4} presents the experimental validation setup and results. Section~\ref{sec5} discusses implementation insights and future deployment considerations. Section~\ref{sec6} concludes the paper and outlines directions for further research.


\begin{table*}[ht]
\centering
\caption{Key Concepts in Multi-Domain 5G Architecture}
\label{tab:MultiDomainConcepts}
\begin{tabular}{@{}p{4cm}p{13cm}@{}}
\toprule
\textbf{Concept} & \textbf{Definition} \\
\midrule
\textbf{Administrative Domain} & \textit{An Administrative Domain is a distinct segment of a 5G network managed by a single MNO, responsible for its own resources, policies, and service operations.}
\\
\midrule
\textbf{Consumer Domain} & \textit{A Consumer Domain is an administrative entity that initiates service requests and specifies application-level requirements within a 5G ecosystem.}. \\
\midrule
\textbf{Provider Domain} & \textit{A Provider Domain represents the administrative authority responsible for offering 5G network, compute, and storage resources.} \\
\midrule
\textbf{Resource Sharing} & \textit{Resource Sharing refers to the controlled exposure and allocation of virtualized network and infrastructure capabilities from the Provider Domain to the Consumer Domain. Rather than sharing physical assets, the Provider offers logical abstractions such as network slices, compute instances, or service APIs through standardized interfaces.} \\
\midrule
\textbf{Inter-Provider Agreement} & \textit{An Inter-Provider Agreement is a formalized understanding established between a Consumer Domain and a Provider Domain to define the terms under which 5G services are provisioned, consumed, and managed.} \\
\midrule
\textbf{Inter-Provider Agreement Lifecycle} & \textit{The process through which domains register, offer, negotiate, establish, and monitor service agreements.} \\
\bottomrule
\end{tabular}
\end{table*}

\begin{table}[ht]
\centering
\caption{Requirements for Inter-Provider Agreements}
\label{tab:requirments}
\begin{tabular}{p{3cm} p{5cm}}
\toprule
\textbf{Requirement} & \textbf{Description} \\
\midrule
Domain Registration & Domains register with verified identity, enabling discovery. \\
\midrule
Advertisement/Discovery & Providers make services visible for others to discover. \\
\midrule
Service Selection & Consumers select services and negotiate agreement terms. \\
\midrule
Privacy-enabled Lifecycle Management & Sensitive information such as SLAs is exchanged privately. \\
\bottomrule
\end{tabular}
\end{table}

\section{Background: Inter-Provider Agreements and Blockchain} \label{sec2} \subsection{Inter-Provider Agreements: Context and Challenges} \label{subsec:InterProviderAgreements}

Building on the concepts introduced in Section~\ref{sec1}, this section formalizes the operational context and technical requirements for inter-provider agreements in multi-domain 5G and emerging 6G environments.

The 5Growth project\footnote{\url{https://5growth.eu/}} exemplifies a collaborative ecosystem in which telecom operators, vertical industries, and third-party service providers coordinate service delivery across heterogeneous administrative domains. As summarized in Table~\ref{tab:MultiDomainConcepts}, each \emph{Administrative Domain} independently manages infrastructure and policies, while \emph{Consumer Domains} initiate service requests and \emph{Provider Domains} offer virtualized resources. By abstracting physical capabilities into logical units—such as network slices or service APIs—providers can expose services without compromising proprietary details.

Within frameworks like 5Growth, automated orchestration and monitoring components streamline \emph{Resource Sharing} and SLA enforcement. Nonetheless, explicit \emph{Inter-Provider Agreements} remain critical to formally govern registration, negotiation, and monitoring processes across domains, as outlined in the \emph{Inter-Provider Agreement Lifecycle} (Table~\ref{tab:MultiDomainConcepts}).

Table~\ref{tab:requirments} details four core requirements for establishing and managing such agreements: verifiable \emph{Domain Registration}, dynamic \emph{Advertisement/Discovery} of available services, consumer-driven \emph{Service Selection}, and \emph{Privacy-enabled Lifecycle Management} to ensure that sensitive business data—such as SLA terms or usage metrics—remain protected. These requirements become increasingly demanding as networks transition toward more open and modular 6G architectures, as envisioned by initiatives like CAMARA \cite{camara} and the GSMA Open Gateway framework.

Although many standardization bodies (e.g., ETSI, 3GPP, IETF) and industry groups (e.g., TM Forum, OpenRAN) address aspects of multi-domain orchestration and network slicing, no unified framework fully resolves the challenges of inter-provider trust and interoperability. This fragmentation, combined with heightened performance and security expectations for 6G, underscores the need for robust agreement mechanisms. As discussed in Section~\ref{sec1}, NGMN’s vision of “trustworthiness by design” \cite{NGMN2023} further emphasizes transparency, privacy, reliability, and resilience as key design goals.

Blockchain and Distributed Ledger Technologies (DLTs) are often proposed to enhance trust and automate contract enforcement in such decentralized settings. However, naive on-chain implementations may expose sensitive contractual details or introduce operational overhead. Earlier work on DLT-based resource allocation and auctions demonstrates blockchain’s potential for multi-domain transparency and trust, but also highlights the limitations of purely public approaches.

A more suitable solution involves a \emph{hybrid} architecture that selectively balances openness and confidentiality. Public processes—such as domain registration (R1), service advertisement (R2), and auditable service selection (R3)—can be handled on-chain to encourage wide participation and transparency. In contrast, sensitive contract elements—such as specific SLA clauses and penalty metrics—should be managed privately (R4), using off-chain or privacy-preserving mechanisms. This hybrid strategy addresses the dual need for collaboration and confidentiality, ensuring that future 6G inter-provider agreements achieve both operational efficiency and competitive protection.

\begin{table}[ht]
\centering
\caption{Key Blockchain Concepts and Definitions}
\label{tab:BlockchainConcepts}
\begin{tabular}{p{2.5cm} p{5cm}}
\toprule
\textbf{Concept} & \textbf{Definition} \\
\midrule
Blockchain & \textit{A distributed ledger technology that records transactions in a secure, immutable, and chronological manner across a decentralized network.} \\
\midrule
Smart Contract & \textit{A self-executing program stored on the blockchain that automatically enforces predefined rules and agreements without intermediaries.} \\
\midrule
Blockchain Node & \textit{An entity that participates in a blockchain network by validating, storing, and propagating transactions and blocks.} \\
\midrule
Consensus Mechanism & \textit{The protocol used by blockchain nodes to agree on the validity of transactions and the state of the ledger, ensuring network consistency.} \\
\midrule
Hybrid Blockchain & \textit{A blockchain architecture that combines elements of both public and private blockchains, enabling controlled access while retaining decentralization benefits.} \\
\midrule
Public and Private Transactions & \textit{Public transactions are visible to all network participants, while private transactions are restricted to authorized parties for confidentiality.} \\
\bottomrule
\end{tabular}
\end{table}

\subsection{Blockchain Foundations for Inter-Provider Agreements} \label{subsec:BlockchainFoundations}

Building on the brief introduction of blockchain technologies in Section~\ref{sec1}, this subsection details the structures and standards relevant to inter-provider agreements.

Blockchain organizes data into \emph{blocks}, each containing transactions or other recorded activities. As shown in Table~\ref{tab:BlockchainConcepts}, each block is cryptographically linked to its predecessor, forming a tamper-resistant ledger. Consensus mechanisms validate transactions without relying on centralized authorities, ensuring data integrity across distributed networks.

Depending on the level of access control, blockchain networks are categorized as public, private, or hybrid. Public blockchains (e.g., Ethereum) allow unrestricted participation, whereas private blockchains (e.g., Hyperledger Fabric) restrict access to authorized users. Hybrid blockchains combine both approaches, enabling selective disclosure of information. They support use cases where some data must remain confidential while other elements remain publicly accessible.

Blockchains also incorporate \emph{smart contracts}, which are self-executing programs that enforce predefined agreements without intermediary oversight. Data recorded on the ledger is immutable and verifiable, providing consistency among participants with varying confidentiality requirements. Hybrid systems allow partial data sharing while maintaining auditability, supporting regulatory and operational requirements \cite{6GTrust}.

Several standardization bodies have issued technical guidelines for blockchain deployment in telecommunications. The European Telecommunications Standards Institute (ETSI) has published specifications for permissioned distributed ledgers (PDLs), including technology overviews (GSPDL001), smart contract deployment guidelines (GSPDL011), and Proof of Concept recommendations (GSPDL005). Other reports address blockchain integration with IoT, edge computing, and artificial intelligence (e.g., GSPDL028, GRPDL032). Parallel efforts by ITU-T, IEEE, and CCSA define reference models, taxonomy, and interoperability standards, with a focus on consensus algorithms, confidentiality, and regulatory compliance.

Within this framework, blockchain can support inter-provider agreements by unifying resource sharing, SLA commitments, and contract enforcement under a single auditable structure. Prior research demonstrates blockchain's applicability to infrastructure sharing \cite{faisal2022beat} and service federation \cite{antevski2022federation}. However, many implementations adopt fully permissioned models that do not address the balance between public accessibility for processes such as onboarding and confidentiality for contract-specific details.

Hybrid blockchain architectures address these requirements by combining public processes (e.g., registration, advertisement) with private channels for sensitive contract terms (e.g., SLA clauses, penalties). Selective privacy and role-based access allow immutability and auditability where needed while protecting business information. This approach avoids the complexity of maintaining separate systems for open and private interactions in multi-domain collaborations.

\begin{figure}[t]
   \centering
   \includegraphics[width=9cm]{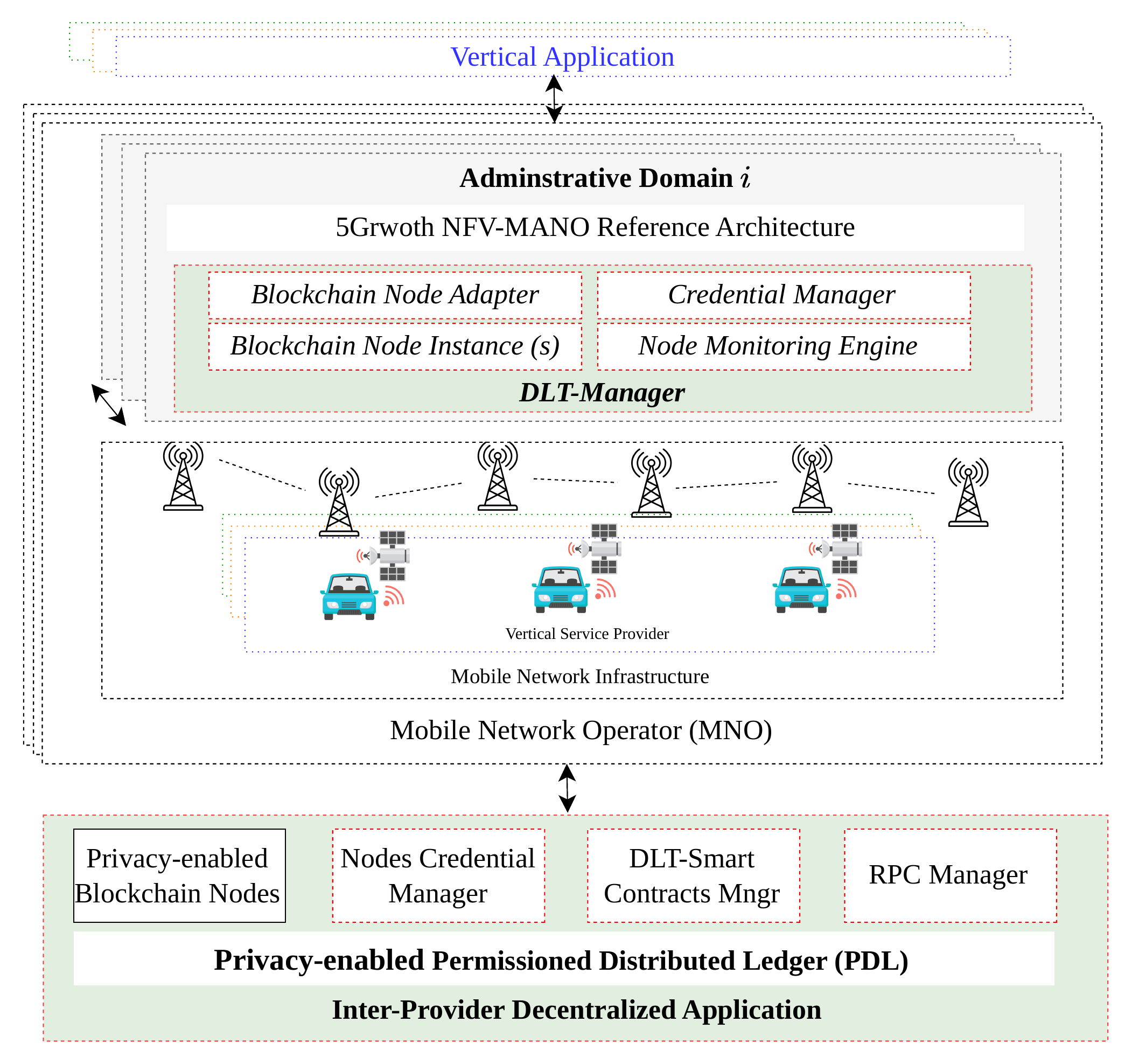}
   \caption{High-level view of the blockchain system architecture for inter-provider agreements integrated with DLT}
   \label{high-level-framework}
\end{figure}

\section{Proposed Hybrid Blockchain Framework for Inter‐Provider Agreements}
\label{sec3}

\begin{figure}[t]
   \centering
   \includegraphics[width=9cm]{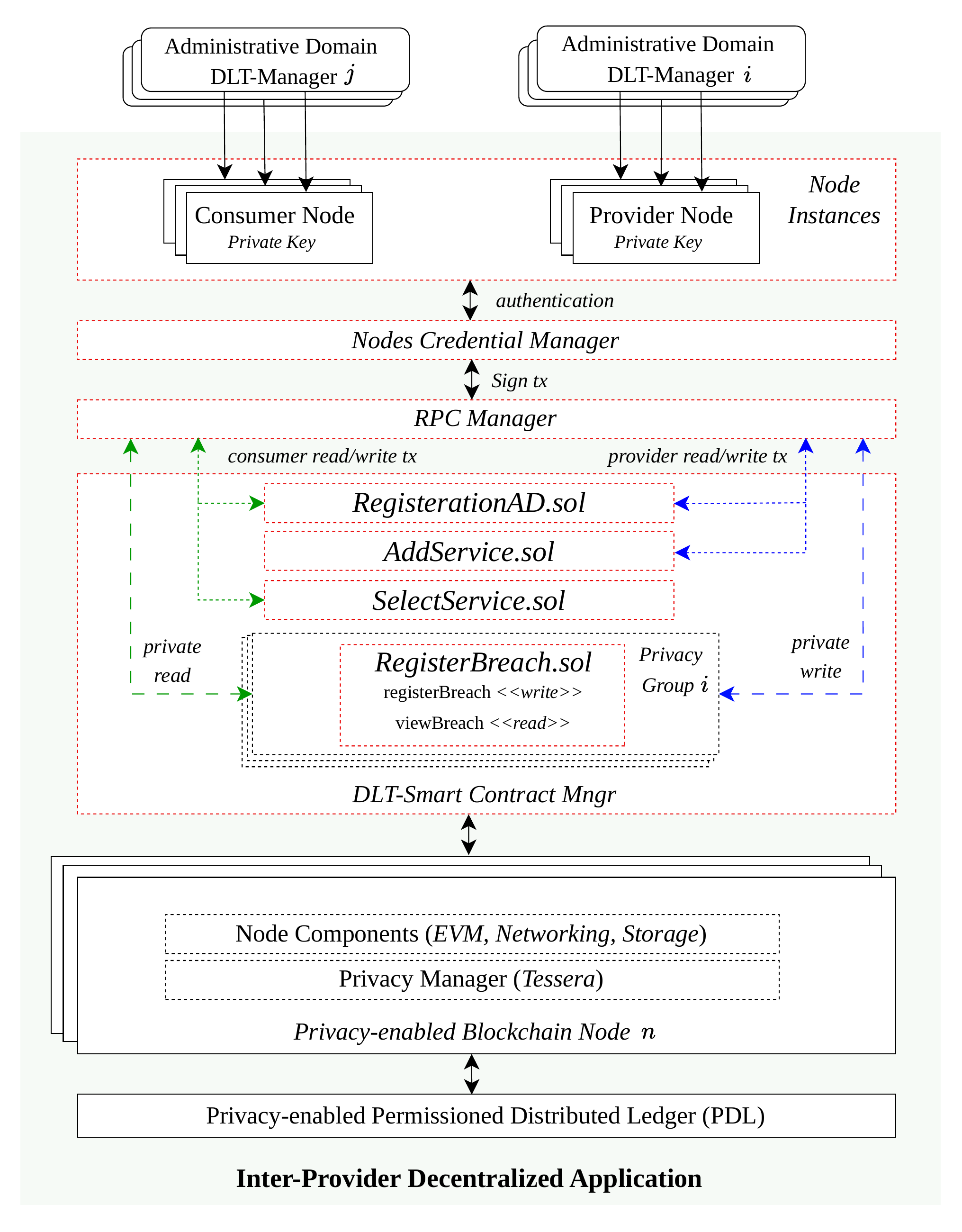}
   \caption{Proposed framework, key components highlighted in color red and their interactions public and private within privacy groups.}
   \label{framework}
\end{figure}

In order to satisfy both open participation and confidentiality in multi-domain orchestration, we propose a hybrid blockchain framework that seamlessly integrates into the 5Growth NFV-MANO reference architecture. As illustrated in Figure~\ref{high-level-framework}, each administrative domain (labeled ``Administrative Domain $i$'') deploys an extended MANO stack, augmented with blockchain-specific components such as a \emph{Blockchain Node Adapter}, a \emph{Credential Manager}, and a \emph{Node Monitoring Engine}. These modules operate under the supervision of a \emph{DLT-Manager} and collectively interface with a shared, privacy-enabled distributed ledger. This ledger embeds decentralized logic within smart contracts (managed by the \emph{DLT-Smart Contracts Mngr}) and is validated by \emph{Blockchain Nodes}, enabling public advertisement of resources while securing confidential SLA details through permissioned transactions. The \emph{RPC Manager} further mediates cross-domain communications, ensuring fine-grained access control and interoperability. This high-level approach (see also Figure~\ref{framework}) reconciles the openness necessary for dynamic resource discovery and on-boarding with the selective privacy critical to safeguarding commercially sensitive terms.

\subsection{Architecture Components}
\label{subsec:architecture_components}
\begin{itemize}
    \item Consumer and Provider Nodes: Each administrative domain deploys at least one node instance—configured with the cryptographic credentials needed to sign and send blockchain transactions. In scenarios where a domain acts solely as a service buyer, it runs a Consumer Node; when it offers services, it uses a Provider Node. In principle, a single domain can operate both node types if it wishes to both consume and provide services.
    \item Node Credential Manager: Acting as the security backbone, the Node Credential Manager stores each node’s private keys and handles critical cryptographic functions (signing transactions, generating address identities, verifying peer certificates, and so forth). When a node issues a request to the DApp, the Node Credential Manager ensures that the correct private key is used for signing, thereby preventing unauthorized or spoofed transactions. This manager also supports role-based access by tying domain credentials to their designated roles (Consumer or Provider).
    \item RPC Manager: Since the blockchain nodes typically expose low-level interfaces, the RPC (Remote Procedure Call) Manager provides a higher-level API for safe and consistent transaction routing. When a node—Consumer or Provider—wants to interact with a smart contract, it communicates with the RPC Manager instead of calling directly into the blockchain node. This design makes it easier to track and validate requests, ensuring correct formatting and permission checks before committing the transaction on-chain. It also simplifies upgrades: changes to the underlying blockchain protocol can be hidden behind the RPC layer without forcing each node to modify its interfaces.
    \item DLT Smart Contract Manager: The “Smart Contract Manager” layer oversees deployment, lifecycle management, and versioning of all relevant contracts in the system. It interacts with both the RPC Manager and the privacy managers to ensure each contract call is routed properly: public calls go to the global ledger state, whereas private calls remain restricted to the specified privacy group.
    \item Privacy-Enabled Blockchain Nodes: Privacy-enabled blockchain implementation, inspired by platforms like Hyperledger Besu. Each node runs an Ethereum Virtual Machine (EVM) for contract execution, networking capabilities for peer discovery and consensus messaging, a storage layer to maintain the global state and chain data, and a dedicated privacy manager (e.g., Tessera) to isolate confidential data from the rest of the network. Thus, while general service announcements and agreements can be processed publicly, sensitive details (e.g., SLA breaches) remain visible only to explicitly authorized privacy groups.
\end{itemize}

\subsection{Smart Contracts and Workflow}
\label{subsec:smart_contracts}

In our design, the DLT Smart Contract Manager orchestrates and maintains four core contracts that map to key steps in the agreement workflow—with the first three handling public or semi‐public functions and the fourth operating within a private group to protect confidential data:

\begin{itemize} 
    \item RegistrationAD.sol: This contract registers each participating administrative domain as either a Consumer or a Provider by storing the domain’s address and role on-chain (step 1), thereby enforcing strict role-based capabilities that ensure Providers can publish services while Consumers are limited to viewing and selecting them; 
    \item AddService.sol: Here, Provider Nodes are enabled to list up to five services on the ledger (step 2) as the contract logs new service entries—including metadata such as price, quality constraints, and availability—ensuring that public service advertisements are immutable, non-duplicative, and transparently accessible; 
    \item SelectService.sol: This contract allows Consumers to choose from the published services by associating the consumer with a specific service in a tamper-proof manner (setp 3), which helps prevent disputes over resource allocation and lays the groundwork for subsequent contract enforcement; 
    \item RegisterBreach.sol (private transactions within privacy group): Operating within a privacy group, this contract records and manages potential SLA breaches, limiting both the registration and viewing of breach data to authorized parties only, thus safeguarding sensitive information while ensuring its immutability and verifiability (setp 4). \end{itemize}

This integrated framework underscores the importance of precise role assignments, transparent service management, secure service selection, and the confidential handling of sensitive breach data, collectively enhancing the reliability and privacy of the multi-domain environment.

\section{Experimental Validation}
\label{sec4}
The simulation environment was deployed on a Linux-based host running Ubuntu 22.04 LTS with a 12th Gen Intel\textsuperscript{\textregistered} Core\texttrademark{} i7-1265U processor (10 cores, up to 4.8~GHz) and 31~GiB of DDR4 memory. The Quorum test network was containerized using Docker and included three Besu nodes and three Tessera nodes to enable privacy-preserving blockchain transactions. The network operated under the IBFT~2.0 consensus protocol, with a validator set of four nodes to ensure Byzantine fault tolerance and fast block finality. An EthSigner proxy facilitated transaction signing, while additional functional components—including a Smart Contract Manager, Credential Manager, and RPC Manager—handled deployment, authentication, and transaction flow. The monitoring stack comprised Prometheus and Grafana, and blockchain activity was visualized using a Block Explorer (Blockscout). The system exposed Web3-compatible JSON-RPC endpoints over both HTTP and WebSocket protocols, supporting decentralized application interactions across administrative domains.

To evaluate the performance of our Quorum-based system, we measured the end-to-end latency of key public and private interactions between consumer and provider nodes. End-to-end latency refers to the total time taken from the initiation of a transaction to its final confirmation on the blockchain. Public interactions—including \texttt{register()}, \texttt{publishService()}, and \texttt{selection()}—were designed to facilitate open service discovery and role assignment. These operations consistently exhibited latencies around 5~seconds reflecting the lower overhead of standard on-chain execution without the need for private payload handling.

In contrast, private operations involving sensitive information were executed within privacy groups to ensure confidentiality between participating parties. A privacy group is a subset of blockchain nodes authorized to view and process private transactions, isolating data from the broader network. Following a successful service selection, a privacy group is dynamically formed between the consumer and the selected provider. Within this group, a dedicated \texttt{RegisterBreach} contract is deployed to record SLA violations. Deploying this private contract incurs higher latency, averaging above 5~seconds, due to the encryption, secure payload distribution, and consensus coordination managed by Tessera. Furthermore, executing the \texttt{registerBreach()} function within the privacy group demonstrates latency variability, ranging from 1 to over 6~seconds. This reflects the dynamic cost of privacy-preserving execution, especially under varying network and synchronization conditions.

These results underscore the performance trade-offs associated with private transactions in Besu. While Tessera enables strict data confidentiality and selective disclosure, it introduces additional computational and communication overhead, particularly for dynamic and sensitive operations. This highlights a key design consideration for decentralized applications operating across administrative domains: balancing transparency and trust with the cost of privacy enforcement.

Furthermore, in our Besu setup using IBFT~2.0, blocks are generated at regular intervals, typically around five seconds. This is consistent with the default configuration in Besu, where the consensus mechanism ensures timely and predictable block production. During our evaluation, we observed that block times remained relatively stable with only minor fluctuations, reflecting the reliability of the IBFT protocol in controlled test environments. While block production timing is relevant for public transactions, it has limited impact on private transaction latency, as private data is exchanged off-chain via Tessera. Off-chain communication refers to processes where data is transferred or verified outside the main blockchain ledger, typically to reduce on-chain computation or preserve confidentiality. Therefore, in privacy-enabled interactions, the main contributors to latency are the encryption, secure communication, and payload handling processes, rather than the block creation interval itself.

\begin{figure}[t]
   \centering
   \includegraphics[width=9cm]{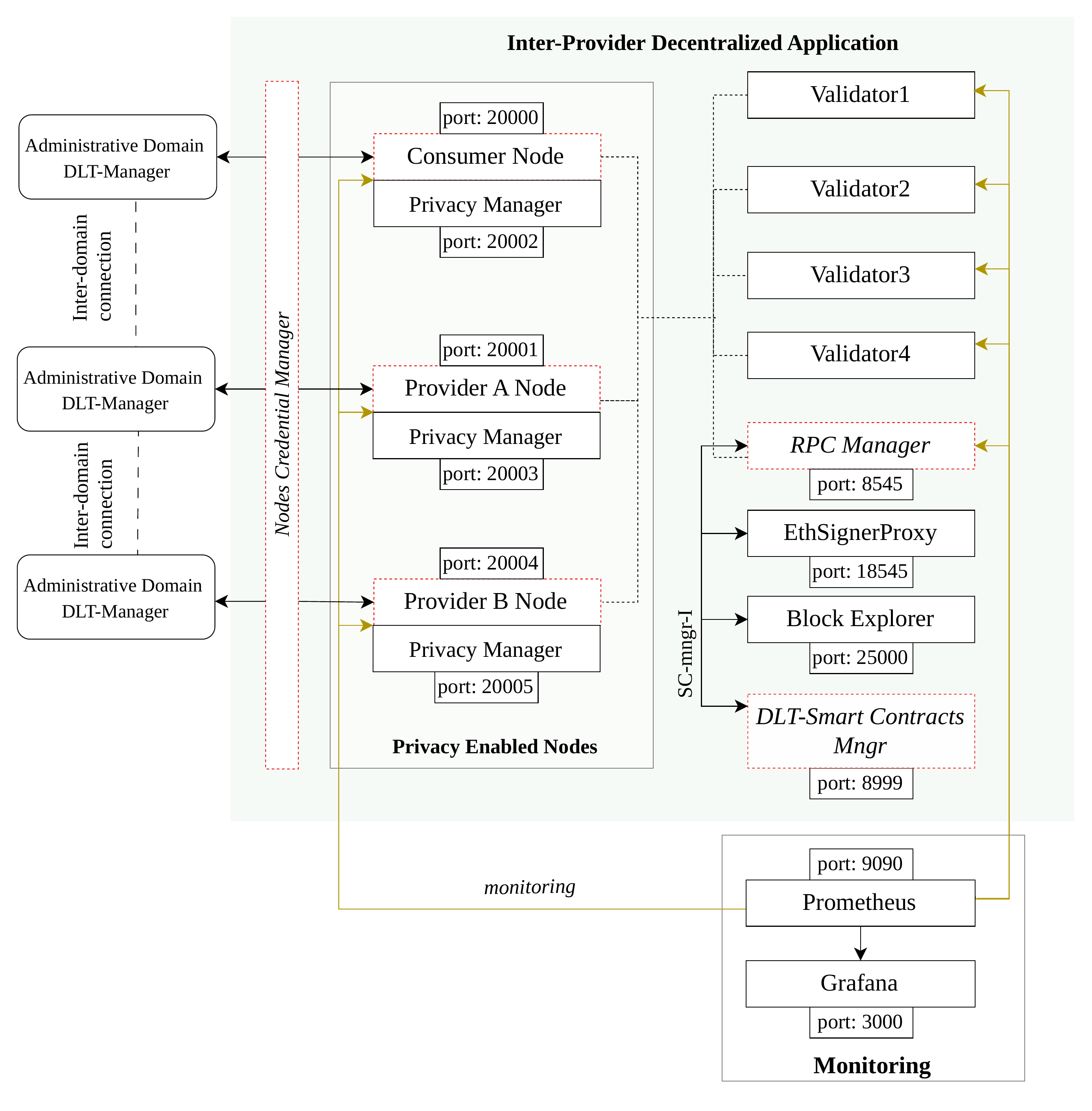}
   \caption{A multi-node Hyperledger Besu network supporting both public and private transactions for Privacy-Enabled Inter-Provider DApp.}
   \label{fig:NodeSetup}
\end{figure}

\begin{table}[tp!]
\centering
\caption{Simulation environment setup}
\begin{tabular}{p{2.5cm} p{5cm}}
\toprule
\textbf{Host OS} & Linux (Ubuntu 22.04 LTS) \\
\midrule
\textbf{CPU} & 12th Gen Intel(R) Core(TM) i7-1265U, 10 cores, max frequency 4.8 GHz \\
\midrule
\textbf{Memory} & 31 GiB DDR4 \\
\midrule
\textbf{Docker Containers} & 
Quorum Members (3 Besu nodes, 3 Tessera nodes) \newline
Validators (4) \newline
EthSigner Proxy \newline
Smart Contract Manager (DLT-SCMgr) \newline
RPC Manager \newline
Credential Manager \newline
Monitoring stack: Prometheus, Grafana \newline
Explorer: Block Explorer (Blockscout) \\
\midrule
\textbf{Quorum Setup} & 
Besu (v23.x), Tessera (v23.4.0) \newline
IBFT 2.0 consensus \newline
Private transactions via Tessera \newline
Web3 JSON-RPC via HTTP \& WebSocket \\
\bottomrule
\end{tabular}
\label{tab:simulation_setup}
\end{table}


\begin{figure}[t!]
    \centering
    \includegraphics[width=\columnwidth]{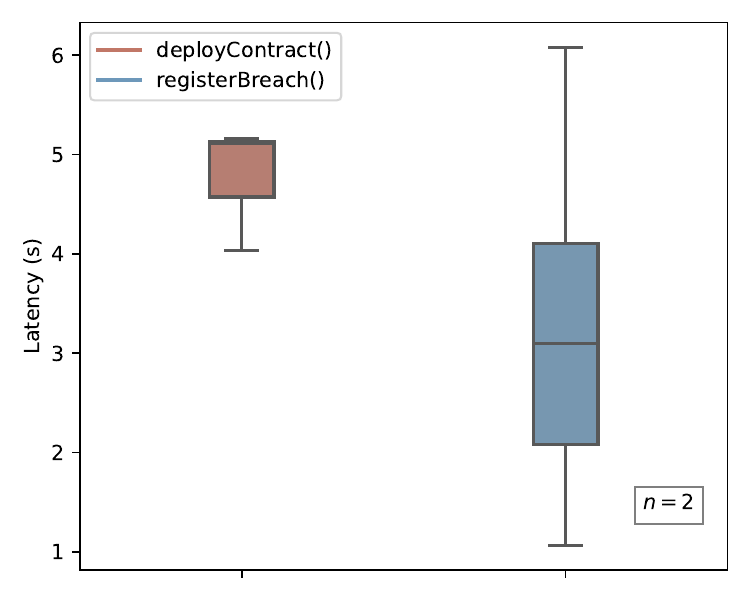}
    \caption{Latency analysis for deploying contract within privacy groups and handling private transactions}
    \label{fig:private_latency}
\end{figure}

\section{Discussion and Lesson Learned}
\label{sec5}
\subsection{Smart Contract Design}
The Proof-of-Concept (PoC) implementation effectively demonstrated how smart contracts can be used to structure inter-provider agreement workflows in a permissioned blockchain setting. The contract design supported key steps such as domain registration, service advertisement, service selection, and SLA breach reporting, with explicit role-based permissions. This approach allowed both transparency in public operations and confidentiality for sensitive interactions, which are essential requirements in multi-domain orchestration scenarios.

Smart contracts deployed on Hyperledger Besu follow the Ethereum execution model, where each transaction is processed by all validating nodes, and each function consumes a certain amount of gas based on its complexity. In our setup, the gas price was set to zero to reflect the economics of a permissioned network, but gas usage was still relevant, as it limited how many transactions could be included per block. High-gas functions such as those that involve storage operations or iterate over data structures had a direct impact on throughput, even without a fee.

We also observed that interactions involving private data—particularly SLA breach reporting—required additional steps beyond standard execution. In these cases, transactions were routed through Tessera, the privacy manager that encrypts data, distributes it securely to authorized participants (privacy group members), and maintains a separate private state. These additional steps introduced measurable delays, especially when multiple parties were involved or network synchronization was affected. As explained in Section~\ref{sec4}, latency for private transactions ranged from 1 to 6 seconds, compared to a more stable ~5 seconds for public ones.

To reduce execution overhead and latency, smart contracts in such environments should be carefully structured. In our system, three contracts handled public logic: RegistrationAD.sol, AddService.sol, and SelectService.sol. A fourth contract, RegisterBreach.sol, was used only within privacy groups to record sensitive information about SLA violations. This separation allowed most interactions to proceed efficiently on the public ledger while preserving confidentiality where needed.

Design patterns that improve efficiency include using hashed references (e.g., storing a hash of an SLA instead of the full terms), emitting encrypted events instead of storing logs on-chain, and avoiding dynamic loops in private contracts. Role-based access should be enforced explicitly at the beginning of each function using conditional checks. For example, providers should only be allowed to publish services, while consumers can only select them. This ensures correctness while simplifying auditing and validation.

Future applications in 6G network environments—where domains may form temporary, dynamic relationships—will benefit from smart contracts that support time-limited access, flexible role changes, and automated state cleanup. Privacy groups should be created when needed and dismantled after agreement expiration. Off-chain authorization mechanisms, such as EIP-712 typed data signatures, can be used to reduce on-chain overhead for common tasks. Structured event logging, rather than storing every detail on-chain, can enable external monitoring tools to reconstruct agreement histories without bloating the ledger.

\subsection{Consensus Mechanism}
The initial PoC was designed to validate the feasibility of executing public and private inter-provider interactions. The chosen consensus configuration, validator setup, and deployment model supports core agreement workflows and observe the behavior of the network under baseline conditions. 
As mentioned in Table \ref{tab:simulation_setup} the consensus mechanism used in our permissioned blockchain setup was IBFT 2.0, which is designed for networks with known and pre-authorized validator nodes. Block production intervals were observed to be approximately five seconds under normal conditions. This timing was consistent across public and private transactions. However, we observed that IBFT block timing had limited influence on the latency of private transactions, which were more affected by Tessera’s off-chain communication than by block interval (see Figure \ref{fig:private_latency}). In contrast, public transactions were finalized according to the IBFT schedule, and their latency was closely aligned with block interval configuration.

The IBFT 2.0 mechanism provides immediate block finality once consensus is reached, eliminating the need for probabilistic confirmation. This is beneficial for inter-provider coordination scenarios, where deterministic agreement on state is required. However, IBFT introduces coordination overhead proportional to the number of validators. As the number of validators increases, the consensus rounds become longer due to the additional communication required to reach supermajority agreement. This impacts the block propagation delay and transaction throughput. Increasing the validator count beyond four can also led to observable increases in average block finalization time. To mitigate this, future deployments should carefully balance validator count against network scalability requirements. For static inter-domain agreements, a smaller validator set may be sufficient. For more dynamic environments with frequent membership changes, a validator selection mechanism based on domain trust levels or operational load may be required.

IBFT 2.0 also relies on stable proposer rotation, and proposer faults can delay block production. To reduce proposer downtime risk, validators should run with high availability configurations, and proposer history should be monitored to identify and exclude persistently slow or faulty nodes. Additionally, IBFT does not support validator churn without manual reconfiguration. For 6G networks where administrative domains may join or leave frequently, dynamic validator management becomes a critical requirement. Current implementations require restarting nodes or editing static configuration files to change the validator set. This limits flexibility. Future work should consider integrating smart contract–based validator voting systems to enable controlled validator updates without network downtime.

\subsection{Improving Blockchain Network Scalability}
The implemented PoC setup validates the end-to-end workflow, measure baseline latency, and evaluate system behavior under controlled conditions. However, scalability remains a key consideration in the design of hybrid blockchain architectures for inter-provider agreement systems, particularly in the context of 6G networks where administrative domains may dynamically join, leave, or form short-term service-level agreements. In such settings, the blockchain infrastructure must support increasing transaction volumes and participant diversity without compromising latency, confidentiality, or state consistency. Hyperledger Besu, as an EVM-compatible platform supporting both public and private transactions, provides mechanisms that can be leveraged to address these challenges. For public interactions—such as service advertisement and selection—transaction batching, load-balanced RPC endpoints, and block parameter tuning (e.g., block interval and gas limit) can improve throughput and responsiveness. For private interactions—such as SLA breach registration or confidential agreement terms—reusing privacy groups across similar workflows, minimizing contract complexity within private state, and deferring non-critical computation to the application layer can reduce coordination and encryption overhead.

In the specific context of inter-provider coordination, scalability should be considered across three dimensions: the number of domains, the frequency of transactions, and the complexity of private workflows. For example, when many domains simultaneously advertise services, a lightweight public contract architecture combined with periodic state compression (e.g., using Merkle root commitments) can reduce ledger growth. In private interactions, if breach events are frequent, a batching mechanism that aggregates and verifies breach records off-chain before committing a summary hash on-chain can preserve confidentiality while reducing transaction load. Similarly, public-private contract separation should be maintained to allow the majority of operations to proceed without entering privacy groups unless strictly necessary.

These patterns can be implemented within the Besu framework without changes to its core protocol, aligning with the requirements for hybrid trust architectures in 6G. The design considerations outlined here reflect a broader need to ensure that blockchain-based coordination mechanisms remain performant as network scale and interaction complexity increase. In future deployments, such scalability patterns can be integrated into orchestration logic and policy frameworks that govern inter-provider interactions, supporting both predictable agreement enforcement and dynamic federation in next-generation network environments.

\section{Conclusion}
\label{sec6}
This article presented a privacy-enabled hybrid blockchain framework for managing inter-provider agreements in 6G networks, with a focus on balancing openness and confidentiality across administrative domains. Built using Hyperledger Besu, the system integrates public and private transaction workflows and enables decentralized service registration, selection, and SLA breach recording. Through a proof-of-concept implementation aligned with the 5Growth architecture, we demonstrated the feasibility of role-based smart contracts and privacy groups for supporting trustworthy and auditable service coordination.

Our evaluation showed that public transactions maintain predictable latency, while private transactions incur additional overhead due to off-chain encryption and synchronization, rather than block production itself. These findings emphasize the importance of careful smart contract design, modular privacy group usage, and minimal on-chain state in private workflows. In addition, IBFT 2.0 consensus behavior highlighted the need to manage validator participation and proposer reliability for scalable and resilient deployment.

Our work provides a foundation for such efforts, offering practical insights into the design of transparent, efficient, and secure agreement systems for next-generation networks. As 6G systems evolve to include increasingly dynamic and federated domains, our future work will explore automated validator onboarding, adaptive privacy group management, and integration of lightweight off-chain verification mechanisms.

\vspace{-1em}
\section*{}

\begin{thebibliography}{10}
\providecommand{\url}[1]{#1}
\csname url@samestyle\endcsname
\providecommand{\newblock}{\relax}
\providecommand{\bibinfo}[2]{#2}
\providecommand{\BIBentrySTDinterwordspacing}{\spaceskip=0pt\relax}
\providecommand{\BIBentryALTinterwordstretchfactor}{4}
\providecommand{\BIBentryALTinterwordspacing}{\spaceskip=\fontdimen2\font plus
\BIBentryALTinterwordstretchfactor\fontdimen3\font minus \fontdimen4\font\relax}
\providecommand{\BIBforeignlanguage}[2]{{%
\expandafter\ifx\csname l@#1\endcsname\relax
\typeout{** WARNING: IEEEtran.bst: No hyphenation pattern has been}%
\typeout{** loaded for the language `#1'. Using the pattern for}%
\typeout{** the default language instead.}%
\else
\language=\csname l@#1\endcsname
\fi
#2}}
\providecommand{\BIBdecl}{\relax}
\BIBdecl

\bibitem{camara}
\BIBentryALTinterwordspacing
GSMA, ``Camara: Telco global api alliance,'' https://www.gsma.com/, accessed: 20-06-2023. [Online]. Available: \url{https://www.gsma.com/futurenetworks/ip_services/understanding-5g/camara-telco-global-api-alliance/}
\BIBentrySTDinterwordspacing

\bibitem{ETSIGSPDL024}
{European Telecommunications Standards Institute}, ``Architecture enhancements for {PDL} service provisioning in telecom networks,'' European Telecommunications Standards Institute, Sophia Antipolis, France, ETSI Group Specification ETSI GS PDL 024 V1.1.1, November 2024, available at: \url{https://www.etsi.org/deliver/etsi_gs/PDL/001_099/024/01.01.01_60/gs_PDL024v010101p.pdf}.

\bibitem{3GPPTS23.251}
3GPP, ``{Technical Specification 23.251: Network Sharing; Architecture and Functional Description},'' 3rd Generation Partnership Project (3GPP), Tech. Rep. TS 23.251, 2022, \url{https://www.3gpp.org/DynaReport/23251.htm}.

\bibitem{ETSI2020Applications}
\BIBentryALTinterwordspacing
{ETSI}, ``{Permissioned Distributed Ledger (PDL); Application Scenarios},'' {European Telecommunications Standards Institute (ETSI)}, Tech. Rep. GR PDL 003 V1.1.1, Dec. 2020. [Online]. Available: \url{https://www.etsi.org/deliver/etsi_gr/PDL/001_099/003/01.01.01_60/gr_PDL003v010101p.pdf}
\BIBentrySTDinterwordspacing

\bibitem{NGMN2023}
{NGMN Alliance}, ``6g trustworthiness considerations,'' \url{https://www.ngmn.org}, October 2023, accessed: 2025-03-10.

\bibitem{TMForumCatalystExample}
``Exploring blockchain-based settlement through catalyst projects,'' TM Forum, 2021, placeholder reference. See \url{https://www.tmforum.org/catalysts/} for various Catalyst projects.

\bibitem{ETSI2019PDLInterop}
{ETSI}, ``Permissioned distributed ledgers (pdl); interoperability and security,'' European Telecommunications Standards Institute (ETSI), Tech. Rep. GR PDL 007, 2019, \url{https://www.etsi.org/deliver/etsi_gr/PDL/001_099/007/01.01.01_60/gr_pdl007v010101p.pdf}.

\bibitem{ETSI2021SmartContracts}
------, ``{Permissioned Distributed Ledgers (PDL); Smart Contracts System Architecture and Functional Specification},'' {European Telecommunications Standards Institute (ETSI)}, Tech. Rep. GR PDL 004 V1.1.1, Feb. 2021.

\bibitem{faisal2022beat}
T.~Faisal, M.~Dohler, S.~Mangiante, and D.~R. Lopez, ``Beat: Blockchain-enabled accountable and transparent network sharing in 6g,'' \emph{IEEE Communications Magazine}, vol.~60, no.~4, pp. 52--56, 2022.

\bibitem{antevski2022federation}
K.~Antevski and C.~J. Bernardos, ``Federation in dynamic environments: Can blockchain be the solution?'' \emph{IEEE Communications Magazine}, vol.~60, no.~2, pp. 32--38, 2022.

\bibitem{xevgenis2025blockchain}
M.~Xevgenis, D.~G. Kogias, and H.~C. Leligou, ``Blockchain in next generation networks (ngns),'' in \emph{Handbook of Blockchain Technology}.\hskip 1em plus 0.5em minus 0.4em\relax Edward Elgar Publishing, 2025, pp. 329--348.

\bibitem{zahir2024performance}
A.~Zahir, M.~Groshev, K.~Antevski, C.~J.~Bernardos, C.~Ayimba, and A.~De~La~Oliva, ``Performance evaluation of private and public blockchains for multi-cloud service federation,'' in \emph{Proceedings of the 25th International Conference on Distributed Computing and Networking}, 2024, pp. 217--221.

\bibitem{wilhelmi2022end}
F.~Wilhelmi, S.~Barrachina-Mu{\~n}oz, and P.~Dini, ``End-to-end latency analysis and optimal block size of proof-of-work blockchain applications,'' \emph{IEEE Communications Letters}, vol.~26, no.~10, pp. 2332--2335, 2022.

\bibitem{afraz2023blockchain}
N.~Afraz, F.~Wilhelmi, H.~Ahmadi, and M.~Ruffini, ``Blockchain and smart contracts for telecommunications: Requirements vs. cost analysis,'' \emph{IEEE Access}, 2023.

\bibitem{antevski2023applying}
K.~Antevski and C.~J. Bernardos, ``Applying blockchain consensus mechanisms to network service federation: Analysis and performance evaluation,'' \emph{Computer Networks}, vol. 234, p. 109913, 2023.

\bibitem{suetor2025overview}
C.~G. Suetor, D.~Scrimieri, A.~Qureshi, and I.-U. Awan, ``An overview of distributed firewalls and controllers intended for mobile cloud computing,'' \emph{Applied Sciences}, vol.~15, no.~4, p. 1931, 2025.

\bibitem{javed2022blockchainglobecom}
F.~Javed and J.~Mangues-Bafalluy, ``Blockchain-based 6g inter-provider agreements: Auction vs. marketplace,'' in \emph{GLOBECOM 2022-2022 IEEE Global Communications Conference}.\hskip 1em plus 0.5em minus 0.4em\relax IEEE, 2022, pp. 1271--1277.

\bibitem{JavedNFVworkshop}
------, ``Blockchain-based sla management for inter-provider agreements,'' in \emph{2022 IEEE Conference on Network Function Virtualization and Software Defined Networks (NFV-SDN)}.\hskip 1em plus 0.5em minus 0.4em\relax IEEE, 2022, pp. 155--161.

\bibitem{javed2025empiricalsmartcontractslatency}
\BIBentryALTinterwordspacing
------, ``An empirical smart contracts latency analysis on ethereum blockchain for trustworthy inter-provider agreements,'' 2025. [Online]. Available: \url{https://arxiv.org/abs/2503.01397}
\BIBentrySTDinterwordspacing

\bibitem{6GTrust}
B.~Veith, D.~Krummacker, and H.~D. Schotten, ``The road to trustworthy 6g: A survey on trust anchor technologies,'' \emph{IEEE Open Journal of the Communications Society}, vol.~4, pp. 581--595, 2023.

\end{thebibliography}


\end{document}